\documentstyle[multicol,aps,epsfig]{revtex}

\begin{document}

\title{Optimizing Epochal Evolutionary Search:\\
Population-Size Independent Theory}
\author{Erik van Nimwegen and James P. Crutchfield\\
Santa Fe Institute, 1399 Hyde Park Road, Santa Fe, NM 87501\\
\{erik,chaos\}@santafe.edu
}

\maketitle


\begin{abstract}
Epochal dynamics, in which long periods of stasis in population
fitness are punctuated by sudden innovations, is a common behavior
in both natural and artificial evolutionary processes. We use a
recent quantitative mathematical analysis of epochal evolution to
estimate, as a function of population size and mutation rate, the
average number of fitness function evaluations to reach the global
optimum. This is then used to derive estimates of and bounds on
evolutionary parameters that minimize search effort.

\begin{center}
Santa Fe Institute Working Paper 98-06-046
\end{center}
\end{abstract}


\begin{multicols}{2}

\tableofcontents


\section{Engineering Evolutionary Search}

Evolutionary search refers to a class of stochastic optimization
techniques---loosely based on processes believed to operate in
biological evolution---that have been applied successfully to a
variety of different problems; see, for example, Refs.
\cite{Back96a,ICGA91,Cham95a,Davis91a,ICGA95,ICGA93,Koza93a} and
references therein. Unfortunately, the mechanisms constraining and
driving the dynamics of evolutionary search on a given problem are not
well understood. In mathematical terms evolutionary search algorithms
are nonlinear population-based stochastic dynamical systems. The
complicated dynamics exhibited by such systems has been appreciated for
decades in the field of mathematical population genetics. For example,
the effects on evolutionary behavior of the rate of genetic variation,
the population size, and the function to be optimized typically cannot
be analyzed separately; there are strong, nonlinear interactions between
them. These complications make an empirical approach to the question of
whether and how to use evolutionary search problematic. The lack of a
unified theory has rendered the literature largely anecdotal and of
limited generality. The work presented here continues an attempt to
unify and extend theoretical work that has been done in the areas of
evolutionary search theory, molecular evolution theory, and mathematical
population genetics. The goal is to obtain a more general and
quantitative understanding of the emergent mechanisms that control the
dynamics of evolutionary search and other population-based dynamical
systems.

Our approach takes a structural view of the search space and solves the
population dynamics as it is constrained by a general architecture for
epochal evolution---a class of population dynamics in which long periods
of stasis are punctuated by rapid innovations. Based on a phenotypically
induced decomposition, the genome space is broken into strongly and
weakly connected sets. From this we motivate several simplifying
assumptions that lead to the class of fitness functions and genetic
operators we analyze. Stated in the simplest possible terms, all of
the resulting population dynamical behavior derives from the interplay
of the architecture, the infinite-population nonlinear dynamics, and 
the stochasticity arising from finite-population sampling.

One might object that important details of real biological evolution,
on the one hand, or of alternative evolutionary search algorithms,
on the other hand, are not described by the resulting class of
evolutionary dynamical systems. Our response is simple: One must start
somewhere. The bottom line is that the results and their predictive
power justify the approach. Moreover, along the way we come to
appreciate a number of fundamental trade-offs and basic mechanisms
that drive and inhibit evolutionary search.

Our results show that a detailed dynamical understanding, announced
in Ref. \cite{Nimw97a} and expanded in Ref. \cite{Nimw97b}, can be
turned to a very practical advantage. Specifically, we determine how
to set population size and mutation rate to reach, in the fewest steps,
the global optimum in a wide class of fitness functions. In other
words, our objective is to minimize the total number of fitness
function evaluations as a function of evolutionary search parameters.

Our analysis provides several insights that are useful knowledge for
engineers even in much more complicated optimization problems (and,
for that matter, for the theory of evolutionary dynamics in biology).
Using these, in a sequel we draw some conclusions about those
optimization problems for which population-based search methods, such as
genetic algorithms and genetic programming (to mention only two
examples), are appropriate.


\section{Landscape Architecture}

The fitness functions characteristic of problems that evolutionary
search or (say) simulated annealing are being used for in practice
are very complicated, almost by definition. On the one hand, detailed
knowledge of the fitness function implies that one does not need
to run an optimization method to find high fitness solutions. On the
other hand, assuming no structure at all leads to a completely random
fitness function for which it is known that {\em any} optimization
algorithm performs as well on average as random search \cite{Wolp95a}.
Not surprisingly, reality is a middle ground.

Therefore, our strategy to understand the workings of evolutionary
search algorithms is to assume some structure in the fitness function
that is germane to search population dynamics and to assume that,
beyond this, there is no other structure. That is, apart from the
structure we impose, the fitness function is as unstructured as can be.

There is a concomitant and compelling biological motivation for our
choice of architecture. This is the common occurrence in natural
evolutionary systems of ``punctuated equilibria''---a process first
introduced to describe sudden morphological changes in the
paleontological record \cite{Gould&Eldredge77}. Similar behavior has
been recently studied experimentally in bacterial colonies
\cite{Elena&Cooper&Lenski96} and in simulations of the evolution of
transfer-RNA secondary structure \cite{Font98a}. It has been argued,
moreover, that punctuated dynamics occurs at both genotypic and
phenotypic levels \cite{Bergman&Feldman96}. This class of behavior
appears robust enough to also occur in artificial evolutionary systems,
such as evolving cellular automata
\cite{Crutchfield&Mitchell95a,MitchellEtAl93c} and populations of
competing programs \cite{Adam95a}.

How are we to think of the mechanisms that cause this evolutionary
behavior? The evolutionary biologist Wright introduced the notion of
``adaptive landscapes'' to describe the (local) stochastic adaptation
of populations to themselves and to environmental constraints
\cite{Wright82a}. This geographical metaphor has had a powerful influence
on thinking about natural and artificial evolutionary processes. The
basic picture is that evolutionary dynamics stochastically crawls along
a surface determined, perhaps dynamically, by the fitness of
individuals, moving to peaks and very occasionally hopping across
fitness ``valleys'' to nearby, and hopefully higher fitness, peaks.

More recently, it has been assumed that the typical fitness functions of
combinatorial optimization and biological evolution can be modeled as
``rugged landscapes'' \cite{Kauf87a,Mack89a}. These are functions with
wildly fluctuating fitnesses even at the smallest scales of single-point
mutations. The result is that these ``landscapes'' possess a large
number of local optima.

\begin{figure}[htbp]
\centerline{\epsfig{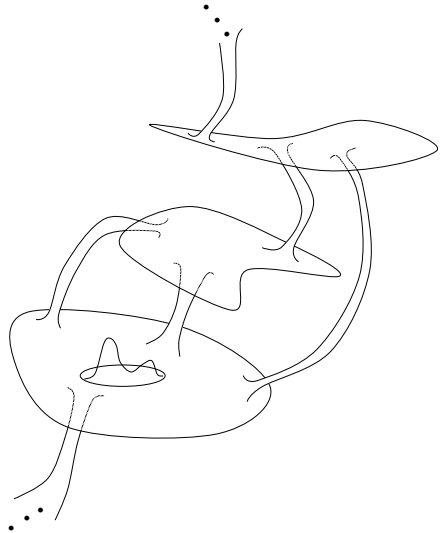}}
\caption{Subbasin and portal architecture underlying
epochal evolution. Roughly speaking, a population diffuses in
the subbasins (large sets) until a portal (a tube) to a higher
fitness subbasin is found.} 
\label{SubbasinPortals}
\end{figure}

At the same time an increasing appreciation has developed, in marked
contrast to this rugged landscape view, that there are substantial
degeneracies in the genotype-to-phenotype and the phenotype-to-fitness
mappings. The crucial role played by these degeneracies has found
important applications in molecular evolution; e.g. see Ref.
\cite{Font93a}. Up to small
fluctuations, when these degeneracies are operating the number of
distinct fitness values in the landscape is often {\em much smaller}
than the number of genotypes. Moreover, due to the high dimensionality
of these genotype spaces, sets of genotypes with approximately equal
fitness tend to form components in genotype space that are connected
by single mutational steps. Finally, due to intrinsic or even
exogenous effects (e.g. environmental) there simply may not exist a
``fitness'' value for each genotype. Fluctuations can induce variation
in fitness such that genotypes with neighboring fitness values are not
distinct at the level of selection. A similar effect can also arise
when a fast rate of change does not allow subtle distinctions in
fitness to become manifest.

When these biological facts are taken into account we end up with an
alternative view to both Wright's ``adaptive'' landscapes and the
more recent ``rugged'' landscapes. That is, the fitness landscape
decomposes into a set of ``neutral networks'' of approximately
isofitness genotypes that are entangled with each other in a complicated
and largely {\em unstructured} fashion; see Fig. \ref{SubbasinPortals}.
Within each neutral network selection is effectively disabled and
neutral evolution dominates. Some of the first steps in understanding
the consequences of neutral evolution (in single neutral networks)
were taken by Kimura in the 1960's using stochastic process analyzes
adopted from statistical physics \cite{Kimurabook}. Despite the early
progress in neutral evolution, a number of fundamental problems remain
\cite{Derrida&Peliti}. Although we will analyze neutral evolution in
the following, we also emphasize the global architectural structure
that connects the neutral networks and drives and constrains epochal
evolutionary search.

\begin{figure}[htbp]
\centerline{\epsfig{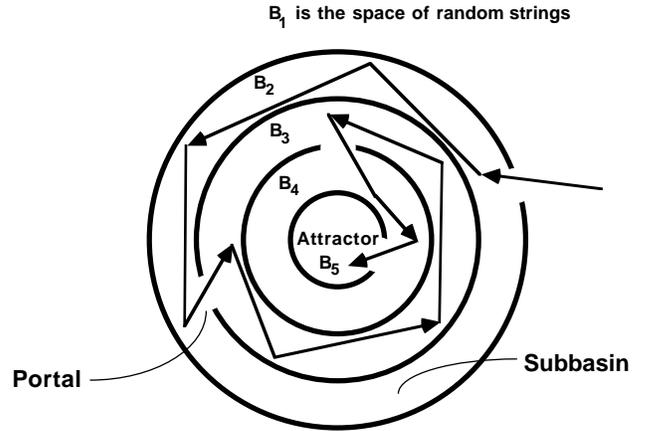}}
\caption{The dimensional hierarchy of subbasins
and portals for the Royal Staircase fitness functions.
} 
\label{DimensionalSubbasinPortals}
\end{figure}

This intuitive view of biologically plausible fitness landscapes---as
a relatively small number of connected neutral nets---is the one that
we adopt in the following analysis. We formalize it by making
several more specific assumptions about the fitness function. First,
we assume that there are $N+1$ different neutral nets, with fitnesses
$1, 2, \ldots, N+1$. Second, we assume that the higher the fitness, the
{\em smaller} the isofitness neutral net volume. That is, there are
fewer strings of high fitness than low. More specifically, we assume
that the proportion of genotype space that is occupied by strings of
fitness $n$ scales as $2^{-K n}$, where $K$ is a measure of the rate at
which the proportion of higher fitness strings decreases with fitness
level. Finally, we assume that strings with fitness $n+1$ can be reached
by a single point mutation from strings with fitness $n$. Specifically,
we assume that the set of strings with fitness $n+1$ is a subspace of
the set of strings with fitness $n$. The resulting architecture is a
modified version of the general subbasin-portal structure of Fig.
\ref{SubbasinPortals} and is illustrated in Fig.
\ref{DimensionalSubbasinPortals}; after Ref. \cite{Crut88e}.

Why assume that strings of higher fitness are nested inside those
of lower fitness? We believe that this assumption is consonant, by
definition, with the very idea of using evolutionary search for
optimization. Imagine, on the contrary, that strings of fitness $n+1$
are more likely to be close to strings with fitness $n-1$ than to
those of fitness $n$. It then seems strange to have selection
preferably replicate strings of fitness $n$ over strings with fitness
$n-1$. One result is that this leads to an increased (ineffective)
search effort centered around strings of fitness $n$. Therefore,
designing a search algorithm to select the current best strings
implicitly assumes that strings of higher fitness tend to be found
close to strings of the current best fitness.

In this way we shift our view away from the geographic metaphor of
evolutionary search ``crawling'' along a fixed and static (smooth or
rugged) ``landscape'' to that of a diffusion process constrained by
the subbasin-portal architecture induced by degeneracies in the
genotype-to-phenotype and phenotype-to-fitness mappings. Moreover, as
will become more apparent, our approach is not just a shift in
architecture, but it also focuses on the dynamics of populations as
they move through the subbasins to find portals to higher fitness.
A side benefit is that it does not limit itself to evolutionary
processes for which a potential function exists; as the landscape
analyses do.

\section{The Royal Staircase Fitness Function}

Under the above assumptions, the class of fitness functions, referred to
as the ``Royal Staircase'', delineated is equivalent to the following
specification:
\begin{enumerate}
\item{Genomes are specified by binary strings
$s = s_1 s_2 \cdots s_L, s_i \in \{0,1\},$ of
length $L = N K$.}
\item{Reading the genome from left to right, the number $I(s)$ of
consecutive $1$s is counted.}
\item{The fitness $f(s)$ of string $s$ with $I$ consecutive ones,
followed by a zero, is $f(s) = 1 + \lfloor I(s) / K \rfloor$. The
fitness is thus an integer between $1$ and $N+1$.}
\item{The (single) global optimum is the genome $s = 1^L$; namely,
the string of all $1$s.}
\end{enumerate}

From this it is easy to see that we have chosen $N$ (consecutive) sets
of $K$ bits to represent the different fitness classes. These sets we
call ``blocks''. The first block consists of the first $K$ bits on
the left, i.e. $s_1 \cdots s_K$. The second block consists of bits
$s_{K+1} \cdots s_{2K}$ and so on. For each of these blocks there is
one ``aligned'' configuration consisting of $K$ $1$s and $2^K-1$
``unaligned'' configurations. If the first block is unaligned,
the string obtains fitness $1$. If the first block is aligned and
the second unaligned, it obtains fitness $2$. If the first two blocks
are aligned and the third unaligned, it obtains fitness $3$, and
so on up to the globally optimal string with all aligned blocks and
fitness $N+1$.

Without affecting the evolutionary dynamics or the underlying
architecture of genotype space, we could have chosen a different
``aligned'' block than the all-$1$s configuration. In fact, we could
have chosen different aligned configurations for the different blocks
and still not affected the dynamics. Furthermore, since we will not be
analyzing
crossover, we could have chosen the locations of the bits of each block
to be anywhere in the genome without affecting the dynamics. The only
constraint, other than the block's ordering, is that we
have $N$ disjoint sets of $K$ bits.

Notice further that the proportion $\rho_n$ of strings with fitness
$n$ is given by:
\begin{equation}
\rho_n = 2^{-K(n-1)} \left( 1- 2^{-K}\right) ~.
\label{nFitStrings}
\end{equation}
The net result is that this fitness function implements the intuitive
idea that increasing fitness is obtained by setting more and more bits
correctly. One can only set correct bit values in sets of $K$ bits at a
time and in blocks from left to right. (Due to the modularity of the
subbasin-portal architecture, and of the resulting theory we present
below, the restriction to uniform block size also could be lifted.) A
genome's fitness is proportional to the number of blocks it has set
correctly. This realizes our view of the underlying architecture as a
set of isofitness genomes that occur in nested neutral networks of
smaller and smaller volume; as shown in Fig.
\ref{DimensionalSubbasinPortals}.

\section{The Genetic Algorithm}

For our analysis of evolutionary search we have chosen a simplified form
of a genetic algorithm (GA) that does not include crossover and that
uses fitness-proportionate selection. The GA is defined by
the following steps.
\begin{enumerate}
\item{Generate a population of $M$ bit strings of length $L = NK$ with
    uniform probability over the space of $L$-bit strings.}
\item{Evaluate the fitness of all strings in the population.}
\item{Stop the algorithm, noting the generation number $t_{\rm opt}$,
if a string with optimal fitness $N+1$ occurs in the population. Else,
proceed.}
\item{Create a new population of $M$ strings by selecting, with
    replacement and in proportion to fitness, strings from the current
    population.}
\item{Mutate each bit in each string of the new population with
    probability $q$.}
\item{Go to step 2.}
\end{enumerate}
The total number $E$ of fitness function queries is $E = M t_{\rm opt}$.
We are interested in the average number $E$ of queries per GA run
required over a large number $R$ of runs. Note that the average total
amount of mutational information introduced into the populations during
a single GA run is $q N K M t_{\rm opt}$.

The main motivation for leaving out crossover is that this greatly
simplifies our analysis. The benefit is that we can make detailed
quantitative predictions of the GA's behavior. Moreover, we believe
that, from the point of view of optimization, the addition of crossover
to the genetic operators only marginally improves the efficiency of the
search. We comment on this, which admittedly is at variance with common
beliefs about evolutionary search, later on. Additional discussion and
supporting evidence can be found in section 6.5 of Ref. \cite{Nimw97b}
and in Ref. \cite{MitchellEtAl93c}.

Notice that our GA effectively has two parameters: the mutation rate
$q$ and the population size $M$. A given search problem is specified
by the fitness function in terms of $N$ and $K$. The central goal of
the following analysis is to find those settings of $M$ and $q$ that
minimize the number $E$ of fitness function queries for given
$N$ and $K$.

\section{Observed Population Dynamics}

The typical dynamics of a population evolving on a landscape of
connected neutral networks, such as defined above, alternates between
long periods of stasis in the population's average fitness (``epochs'')
and sudden increases in the average fitness (``innovations''). (See,
for example, Fig. 1 of Ref. \cite{Nimw97b}.) As was first pointed out
in the context of molecular evolution in Ref.
\cite{Huynen&Stadler&Fontana}, the best individuals in the population
diffuse over the neutral network (``subbasin'') of isofitness genotypes
until one of them discovers a connection (``portal'') to a neutral
network of even higher fitness. The fraction of individuals on this
new network then grows rapidly, reaching an equilibrium after which the
new subset of most-fit individuals diffuses again over the new neutral
network. In addition to the increasing attention paid to this type of
epochal evolution in the theoretical biology community
\cite{Forst&Reidys&Weber95,Huynen95,Newman&Engelhardt98,Reidys98b,Weber97},
recently there has also been an increased interest by evolutionary
search theorists \cite{Barnett97,Haygood}.

The GA just defined is the same as that studied in our earlier analyses
\cite{Nimw97a,Nimw97b}. Also, the Royal Staircase fitness function
defined above is very similar to the ``Royal Road'' fitness function
that we used there. It should not come as a surprise, therefore, that
qualitatively the GA's experimentally observed behavior is very
similar to that reported in Refs. \cite{Nimw97a} and \cite{Nimw97b}.
Moreover, most of the theory developed there for epochal evolutionary
dynamics carries over to the Royal Staircase class of fitness functions.

We now briefly recount the experimentally observed behavior of typical
Royal Staircase GA runs. The reader is referred to Ref. \cite{Nimw97b}
for a detailed discussion of the dynamical regimes this type of GA
exhibits under a range of different parameter settings.

\end{multicols}

\begin{figure}[htbp]
\centerline{\epsfig{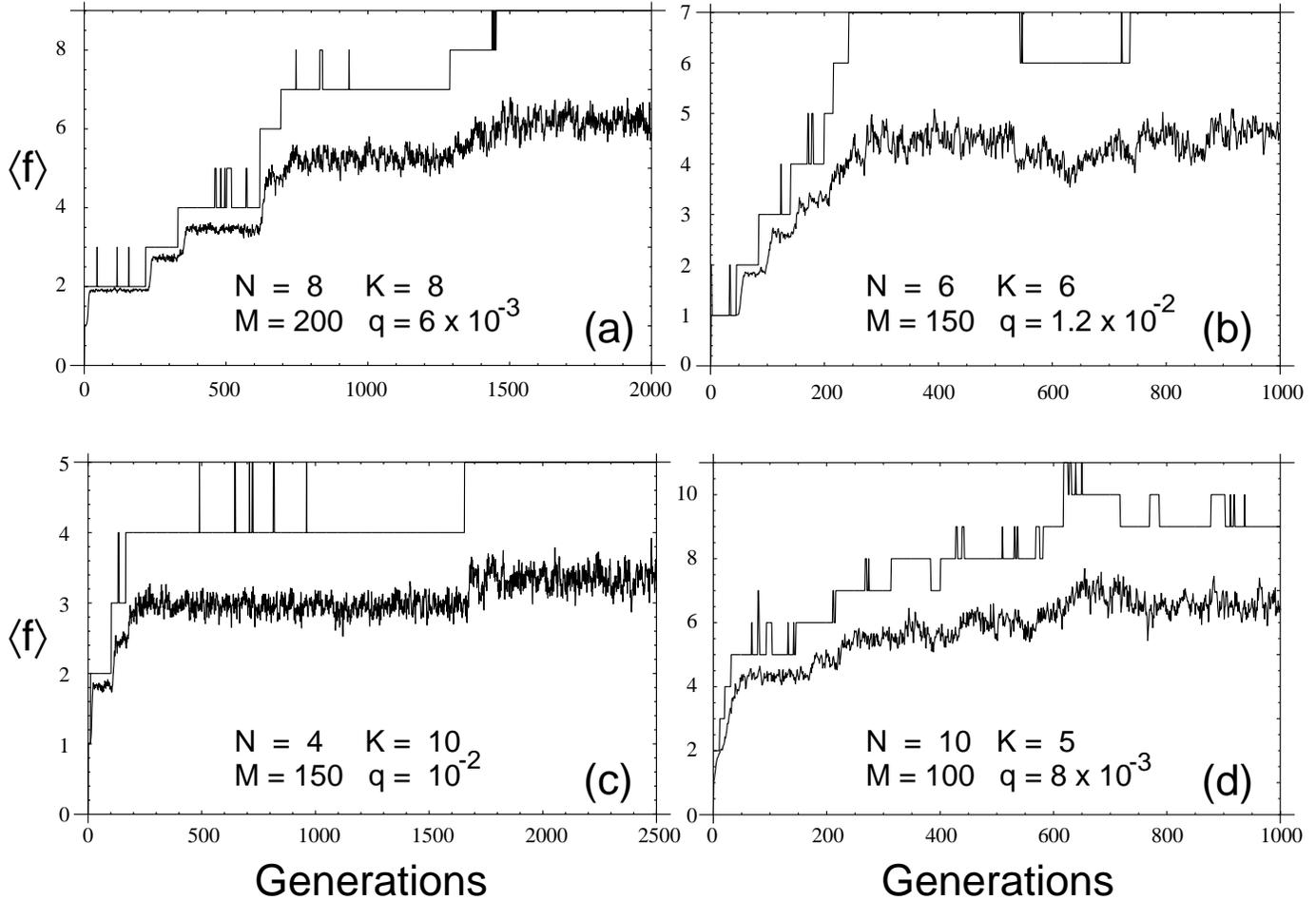}}
\caption{Examples of the Royal Staircase GA population dynamics with
  different parameter settings. The four plots show best fitness in
  the population (upper lines) and average fitness in the population
  (lower lines) as a function of time, measured in generations. The
  fitness function and GA parameters are given in each plot. In each
  case we have chosen $q$ and $M$ in the neighborhood of their optimal
  settings (see later) for each of the four values of $N$ and $K$.} 
\label{fig_runs}
\end{figure}

\begin{multicols}{2}

Figure \ref{fig_runs} shows the GA's behavior with four different
parameter settings. The vertical axes show the best fitness in the
population (upper lines) and the average fitness in the population
(lower lines) as a function of the number of generations. Each figure
is produced from a single GA run. In all of these runs the average
fitness $\langle f \rangle$ in the population goes through stepwise
changes early in the run, alternating epochs of stasis with sudden
innovations in fitness. Later in the run, the average fitness tends
to have higher fluctuations. Notice also that $\langle f \rangle$
roughly tracks the epochal
behavior of the best fitness in the population. Notice, too, that often
the best fitness shows a series of innovations to higher fitness that
are lost. Eventually these innovations ``fixate'' in the population.
Finally, for each of the four settings of $N$ and $K$ we have chosen
the values of $M$ and $q$ such that the total number $E$ of fitness
function evaluations to reach the global optimum for the first time is
roughly minimal. Thus, the four plots illustrate the GA's typical
dynamics close to optimal $(M,q)$-parameter settings---the analysis
for which begins in the next section.

There is a large range, almost a factor of $10$, in times to reach the
global optimum across the runs. Thus, there can be a strong parameter
dependence in search times. Moreover, the {\em variance} of the
total number $E$ of fitness function evaluations is the same size
as the average $E$. Thus, there are large run-to-run variations in the
time to reach the global optimum, even with the parameters held
constant. This is true for all parameter settings with which
we experimented, of which only a few are reported here.

Figure \ref{fig_runs}(a) plots the results of a GA run with $N=8$
blocks of $K=8$ bits each, a mutation rate of $q=0.006$, and a
population size of $M=200$. During the epochs, the best
fitness in the population jumps up and down several times before it
finally jumps up and the new more-fit string stabilizes in the
population. This transition is reflected in the average fitness
also starting to move upward. In this particular run, it took the
GA approximately $3 \times 10^5$ fitness function evaluations to
reach the global optimum for the first time. Over 250 runs the GA
takes on average $5 \times 10^5$ fitness function evaluations to
reach the global optimum for these parameters.
The inherent large per-run variation means in this case that some
runs take less than $10^5$ function evaluations and that others take
many more than $10^6$.

Figure \ref{fig_runs}(b) plots a run with $N=6$ blocks of length $K=6$
bits, a mutation rate of $q=0.012$, and a population size of $M=150$.
The GA reached the global optimum after approximately $3 \times 10^4$
fitness function evaluations. On average, the GA uses approximately
$5 \times 10^4$ fitness function evaluations to reach the global fitness
optimum. Notice that somewhere after generation $500$ the global
optimum is lost again from the population. It turns out that this is a
typical feature of the GA's behavior for parameter settings close to
those that give minimal $E$. The global fitness optimum often only
occurs in relatively short bursts after which it is lost again
from the population. Notice also that there is only a small difference
in $\langle f \rangle$ depending whether the best fitness is either
$6$ or $7$.

Figure \ref{fig_runs}(c) shows a run for a small number ($N=4$) of large
($K=10$) blocks. The mutation rate is $q=0.01$ and the population size
is again $M=150$. As in all three other runs we see that the average
fitness goes through epochs punctuated by rapid increases of
average fitness. We also see that the best fitness in the population
jumps up several times before the population fixates on a higher
fitness. The GA takes about $2 \times 10^5$ fitness function
evaluations on average to reach the global optimum for these parameter
settings. In this run, the GA just happened to have taken about
$2.5 \times 10^5$ fitness function evaluations.

Finally, Fig. \ref{fig_runs}(d) shows a run with a large number
($N=10$) of smaller ($K=5$) blocks. The mutation rate is $q=0.008$ and
the population size is $M=100$. Notice that in this run, the best
fitness in the population alternates several times between fitnesses
$7$, $8$, and $9$ before it reaches the global fitness optimum of $11$.
After it has reached the global optimum for several time steps, the
global optimum disappears again and the best fitness in the population
alternates between $9$ and $10$ from then on. It is notable that this
intermittent behavior of the best fitness is barely discernible in the
behavior of the average fitness. It appears to be lost in the ``noise''
of the average fitness fluctuations. The GA takes about $10^5$ fitness
function evaluations on average at these parameter settings to reach
the global optimum; while in this particular run the GA took only
$6 \times 10^4$ fitness function evaluations.

Again, we stress that there are large fluctuations in the total number
of fitness evaluations to reach the global optimum between runs. One
tentative conclusion is that, if one is only going to use a GA for a
few runs on a specific problem, there is a large range in parameter
space for which the GA's performance is statistically equivalent. On
the one hand, the large fluctuations in the GA's search dynamics make
it hard to predict for a single run how long it is going to take to
reach the global optimum. On the other hand, this prediction is largely
insensitive to the precise parameter settings in a ball centered around
the optimal parameter settings. Thus, the large fluctuations lead to a
large ``sweet spot'' of GA parameters, but the GA does not reliably
discover the global optimum within a fixed number of fitness function
evaluations.

\section{Statistical Dynamics of Evolutionary Search}

In Refs. \cite{Nimw97a} and \cite{Nimw97b} we developed the statistical
dynamics of genetic algorithms to analyze the behavioral regimes of a
GA searching the Royal Road fitness function. The analysis here builds
on those results. Due to the strong similarities we will only briefly
introduce this analytical approach to evolutionary dynamics. The reader
is referred to Ref. \cite{Nimw97b} for an extensive and detailed
exposition. There the reader also will find a review of the alternative
methodologies for GA theory developed by Vose and collaborators
\cite{Nix&Vose91,Vose91,Vose&Liepins91}, by Prugell-Bennett, Rattray,
and Shapiro\cite{PrugelBennett&Shapiro94,PrugelBennett&Shapiro96,Rattray&Shapiro96},
and in mathematical population genetics.

From a microscopic point of view, the state of an evolving population
is only fully described when a list $\cal{S}$ of all genotypes with
their frequencies of occurrence in the population is given. The
evolutionary dynamics is implemented via the conditional transition
probabilities $P(\cal{S'} | \cal{S})$ that the population
at the next generation will be the microscopic state $\cal{S'}$.
For any reasonable genetic representation, there will
be an enormous number of these microscopic states $\cal{S}$ and
their transition probabilities. This makes it almost impossible
to quantitatively study the dynamics at this microscopic level.

More practically, a full description of the dynamics on the
level of microscopic states $\cal{S}$ is neither useful nor typically
of interest. One is much more likely to be concerned with relatively
coarse statistics of the dynamics, such as the evolution of the best
and average fitness in the population or the waiting times for evolution
to produce a string of a certain quality. The result is that
quantitative mathematical analysis faces the task of finding a
description of the evolutionary dynamics that is simple enough to be
tractable numerically or analytically and that, moreover, facilitates
the prediction the quantities of interest to a practitioner.

With these issues in mind, we specify the state of the population at
any time by some relatively small set of ``macroscopic'' variables.
Since this set of variables intentionally ignores vast amounts of
microscopic detail, it is generally impossible to exactly describe
the GA's dynamics in terms of these macroscopic variables. In order
to acheive the benefits of a coarser description,
however, we assume that given the state of the macroscopic
variables, the population has equal probabilities to be in {\em any}
of the microscopic states consistent with the specified state of the
macroscopic variables. This ``maximum entropy'' assumption lies at
the heart of statistical mechanics in physics.

We assume in addition that in the limit of infinite population size,
the resulting equations of motion for the macroscopic variables become
closed. That is, for infinite populations, we assume that we can
exactly predict the state of the macroscopic variables at the next
generation, given the present state of the macroscopic variables. This
limit is analogous to the thermodynamic limit in statistical mechanics;
the corresponding assumption is analogous to ``self-averaging'' of the
dynamics in this limit.

The key, and as yet unspecified step, in developing such a
thermodynamic model of evolutionary dynamics is to find an
appropriate set of macroscopic variables that satisfies the above
assumptions. In practice this is difficult. Ultimately, the suitability
of a set of macroscopic variables has to be verified by comparing
theoretical predictions with experimental measurements. In choosing
such a set of macroscopic variables one is guided by our knowledge of
the fitness function and the search's genetic operators.

To see how this choice is made imagine that strings can have only two
possible values for fitness, $f_A$ and $f_B$. Assume also that under
mutation all strings of type $A$ are equally likely to turn into
type-$B$ strings and that all strings of type $B$ have equal
probability to turn into strings of type $A$. In
this situation, it is easy to see that we can take the macroscopic
variables to be the relative proportions of $A$ strings and $B$ strings
in the population. Any additional microscopic detail, such as the
number of $1$s in the strings, is not required or relevant to the
evolutionary dynamics. Neither selection nor mutation distinguish
different strings within the sets $A$ and $B$ on the level of the
proportions of $A$'s and $B$'s they produce in the next generation.

Similarly, our approach describes the state of the population at any
time only by the distribution of {\em fitness} in the population. That
is, we group strings into equivalence classes of equal fitness and
assume that, on the level of the fitness distribution, the dynamics
treats all strings within a fitness class as equal. At the macroscopic
(fitness) level of the dynamics, we know that a string of fitness $n$
has the first $n-1$ blocks aligned and the $n$th block in one of the
$2^K-1$ other unaligned configurations. The maximum entropy assumption
specifies that for all strings of fitness $n$, the $n$th block is
{\em equally likely} to be in any of the $2^K-1$ unaligned
configurations and that each of the blocks $n+1$ through $N$ are equally
likely to be in any of their possible $2^K$ block configurations.

Various reasons suggest the maximum entropy approximation will not
be valid in practice. For example, the fixation due to finite
population size makes it hard to believe that all unaligned block
configurations in all strings are random and {\em independent}.
For large populations, fortunately, the assumption is compelling.
In fact, in the limit of very large population sizes,
typically $M > 2^L$, the GA's dynamics on the level of fitness
distributions accurately captures the fitness distribution dynamics
found experimentally \cite{Nimw97b}. In any case, we will solve
explicitly for the fitness distribution dynamics in the limit of
infinite populations using our maximum entropy assumption and then
show how this solution can be used to solve for the finite-population
dynamics.

The essence of our statistical dynamics approach then is to describe
the population state at any time during a GA run by a relatively small
number of macroscopic variables---variables that in the limit
of infinite populations self-consistently describe the dynamics
at their own level. After obtaining this infinite population
dynamics explicitly, we then use it to solve for the
GA's dynamical behaviors with finite populations.

Employing the maximum entropy principle and focusing on fitness
distributions is also found in an alternative statistical mechanics
approach to GA dynamics developed by Pru\"gel-Bennett, Rattray, and
Shapiro \cite{PrugelBennett&Shapiro94,PrugelBennett&Shapiro96,Rattray&Shapiro96}.
In their approach, however, maximum entropy is assumed with respect to
the ensemble of entire GA {\em runs}. Specifically, in their analysis
the {\em average} dynamics, averaged over many runs, of the first few
cumulants of the fitness distribution are predicted theoretically.
The result is that almost all of the relevant behavior, e.g. as
illustrated in Fig. \ref{fig_runs} of the previous section, is
averaged away. In contrast,
our statistical dynamics approach applies maximum entropy only
to the population's {\em current} state, given its current fitness
distribution. The result is that for finite populations we do
not assume that the GA dynamics is self-averaging. That is, two runs,
with equal fitness distributions $\vec{P}$ at time $t$, are not assumed
to have the same future macroscopic behavior. They are assumed only to
have the same probability distribution of possible futures.

\subsection{Generation Operator}

The macroscopic state of the population is determined by its fitness
distribution, denoted by a vector
$\vec{P} = \{P_1 , P_2 , \ldots , P_{N+1} \}$, where the components
$0 \leq P_n \leq 1$ are the proportion of individuals in the population
with fitness $n = 1, 2, \ldots, N+1$. We refer to $\vec{P}$ as the
phenotypic quasispecies, following its analog in molecular evolution
theory \cite{Eigen71,Eigen&Schuster77}. Since $\vec{P}$ is a
distribution, we have the normalization condition:
\begin{equation}
\sum_{n=1}^{N+1} P_n = 1.
\label{Normalization}
\end{equation}
The average fitness $\langle f \rangle$ of the population is given by:
\begin{equation}
\langle f \rangle = \sum_{n=1}^{N+1} n P_n.
\end{equation}

In the limit of infinite populations and under our maximum entropy
assumption, we can construct a generation operator ${\bf G}$ that
maps the current fitness distribution $\vec{P}(t)$
deterministically into the fitness distribution $\vec{P}(t+1)$ at the
next time step; that is,
\begin{equation}
\vec{P}(t+1) = {\bf G} [ \vec{P}(t) ] ~.
\end{equation}

The operator ${\bf G}$ consists of a selection operator ${\bf S}$
and a mutation operator ${\bf M}$:
\begin{equation}
{\bf G} = {\bf M} \cdot {\bf S}.
\end{equation}
The selection operator encodes the fitness-level effect of selection
on the population; and the mutation operator, the fitness-level effect
of mutation. Below we construct these operators for our GA and the Royal
Staircase fitness function explicitly. For now we note that the infinite
population dynamics can be obtained by iteratively applying the operator
${\bf G}$ to the initial fitness distribution $\vec{P}(0)$. Thus, the
macroscopic equations of motion are formally given by
\begin{equation}
\vec{P}(t) = {\bf G}^{(t)} [ \vec{P}(0) ] ~.
\end{equation}
Recalling Eq. (\ref{nFitStrings}) it is easy to see that the initial
fitness distribution $\vec{P}(0)$
is given by:
\begin{equation}
\label{init_fit_dist}
P_n(0) = 2^{-K (n-1)} \left( 1- 2^{-K}\right) \;, ~ 1 \leq n \leq N ~,
\end{equation}
and
\begin{equation}
P_{N+1}(0) = 2^{-K N}.
\end{equation}
As shown in Refs. \cite{Nimw97a} and \cite{Nimw97b}, despite
${\bf G}$'s nonlinearity, it can be linearized such that the $t$th
iterate ${\bf G}^{(t)}$ can be directly obtained by solving for the
eigenvalues and eigenvectors of ${\bf G}$.

For very large populations ($M > 2^L$) the dynamics of the fitness
distribution obtained from GA simulation experiments is accurately
predicted by iteration of the operator ${\bf G}$. It is noteworthy,
though, that this ``infinite'' population dynamics is qualitatively
very different from the behavior shown in Fig. \ref{fig_runs}. For
large populations strings of {\em all} fitnesses are present in the
initial population and the average fitness increases smoothly and
monotonically to an asymptote over a small number of generations. 
(This limit is tantamount to an exhaustive search, requiring as it
does ${\cal O} (2^L)$ fitness function evaluations.) Despite this
seeming lack of utility, in the next section we show how to use the
infinite population dynamics given by ${\bf G}$ to describe the
finite-population behavior.

\subsection{Finite Population Dynamics}

There are two important differences between the infinite-population
dynamics and that with finite populations. The first is that with
finite populations the components $P_n$ cannot take on arbitrary values
between $0$ and $1$. Since the number of individuals with fitness $n$ in the
population is necessarily integer, the values of $P_n$ are quantized
to multiples of $1/M$. The space of allowed finite population fitness
distributions thus turns into a regular lattice in $N+1$ dimensions
with a lattice spacing of $1/M$ within the simplex specified by
normalization (Eq. (\ref{Normalization})). Second, the dynamics of
the fitness distribution is no longer deterministic. In general, we
can only determine the conditional probabilities
${\rm Pr} [ \vec{Q} | \vec{P} ]$ that a certain
fitness distribution $\vec{P}$ leads to another $\vec{Q}$ in the next
generation.

Fortunately, the probabilities
${\rm Pr} [ \vec{Q} | \vec{P} ]$ are simply given by a multinomial
distribution with mean ${\bf G} [ \vec{P} ]$, which in
turn is the result of the action of the infinite population dynamics.
This can be understood in the following way. In creating the population
for the next generation individuals are selected, copied, and mutated,
$M$ times from the {\em same} population $\vec{P}$. This means that for
each of the $M$ selections there are equal probabilities that a string
of fitness $n$ will be produced in the next generation. For an infinite
number of selections the final proportions $Q_n$ of strings with fitness
$n$ are just the probabilities to produce a single individual with
fitness $n$ with a single selection. That is, given an infinite
population ${\bf G} [ \vec{P} ]$ we have for a finite population that
the fitness distribution $\vec{Q}$ is a random sample of size $M$ of
the distribution ${\bf G} [ \vec{P} ]$.

Putting these observations together, if we write $Q_n = m_n/M$, with
$0 \leq m_n \leq M$ being integers, we have:
\begin{equation}
{\rm Pr} [ \vec{Q} | \vec{P} ] = M!
\prod_{n=1}^{N+1} \frac{\left({\bf G}_n [ \vec{P} ]
  \right)^{m_n}}{m_n!} ~.
\end{equation}
We see that for any finite-population fitness distribution $\vec{P}$
the operator ${\bf G}$ still gives the GA's {\em average} dynamics
over one time step, since the {\em expected} fitness distribution at
the next time step is ${\bf G} [ \vec{P} ]$. Note that
the components ${\bf G}_n [ \vec{P} ]$ need not be multiples
of $1/M$. Therefore the {\em actual} fitness distribution $\vec{Q}$
at the next time step is not ${\bf G} [ \vec{P} ]$, but is
instead one of the lattice points of the finite-population state space.
Since the variance around the expected distribution
${\bf G} [ \vec{P} ]$ is proportional to $1/M$,
$\vec{Q}$ is likely to be close to
${\bf G} [ \vec{P} ]$.

\subsection{Epochal Dynamics}

For finite populations, the expected change $\langle d\vec{P} \rangle$
in the fitness distribution over one generation is given by:
\begin{equation}
\langle d\vec{P}\rangle = {\bf G} [ \vec{P} ] -\vec{P}.
\end{equation}
Assuming that some component $\langle d P_i \rangle$ is much smaller
than $1/M$, the actual change in component $P_i$ is likely to be
$d P_i = 0$ for a long succession of generations. That is, if the size
of the ``flow'' $\langle dP_i\rangle$ in some direction $i$ is much
smaller than the lattice spacing ($1/M$) for the finite population, we
expect the fitness distribution to not change in direction (fitness)
$i$. In Refs.
\cite{Nimw97a} and \cite{Nimw97b} we showed that this is the mechanism
that causes epochal dynamics for finite populations. More formally,
each epoch $n$ of the dynamics corresponds to the population being
restricted to a region in the $n$-dimensional lower-fitness subspace
of fitnesses $1$ to $n$ of the macroscopic state space. Stasis occurs because
the flow out of this subspace is much smaller than the finite-population
induced lattice spacing.

As Fig. \ref{fig_runs} illustrates, each epoch in
the average fitness is associated with a constant value of the best
fitness in the population. More detailed experiments reveal that not
only is $\langle f \rangle$ constant on average during the epochs, in
fact the entire fitness distribution $\vec{P}$ is constant on
average during the
epochs. We denote by $\vec{P}^n$ the average fitness distribution
during the generations when $n$ is the highest fitness in the
population. As was shown in Ref. \cite{Nimw97b}, each epoch fitness
distribution $\vec{P}^n$ is the unique fixed point of the operator
${\bf G}$ restricted to the $n$-dimensional subspace of strings with
$1 \leq f \leq n$. That is, if ${\bf G}^n$ is the projection of the
operator ${\bf G}$ onto the $n$-dimensional subspace of fitnesses
from $1$ up to $n$, then we have:
\begin{equation}
{\bf G}^n [ \vec{P}^n ] = \vec{P}^n ~.
\end{equation}
The average fitness $f_n$ in epoch $n$ is then given by:
\begin{equation}
f_n = \sum_{j=1}^{n} j P^n_j.
\end{equation}
Thus, the fitness distributions $\vec{P}^n$ during epoch $n$ are
obtained by finding the fixed point of ${\bf G}$ restricted to the
first $n$ dimensions of the fitness distribution space. We will
construct the operators ${\bf G}^n$ explicitly below for our GA and
solve analytically for the epoch fitness distributions $\vec{P}^n$
as a function of $n$, $K$, and $q$.

The global dynamics can be viewed as an incremental discovery of
successively more dimensions of the fitness distribution space. In
most realistic settings, it is
typically the case that population sizes $M$ are much smaller than
$2^L$. Initially, then, only strings of low fitness are present in the
initial population, as can also be seen from Eq. (\ref{init_fit_dist}).
The population then stabilizes on the epoch fitness distribution
$\vec{P}^n$ corresponding to the best fitness $n$ in the initial
population. The fitness distribution fluctuates around $\vec{P}^n$
until a string of fitness $n+1$ is discovered and spreads through the
population. The population then settles into fitness distribution
$\vec{P}^{n+1}$ until a string of fitness $n+2$ is discovered, and so
on, until the global optimum at fitness $N+1$ is found. In this way,
the global dynamics can be seen as stochastically hopping between the
different epoch distributions $\vec{P}^n$.

Whenever mutation creates a string of fitness $n+1$, this string may
either disappear before it spreads (seen as the isolated jumps in best
fitness in Fig. \ref{fig_runs}) or it may spread, leading the
population to fitness distribution $\vec{P}^{n+1}$. We call the latter
process an innovation. Fig. \ref{fig_runs} also showed that it is
possible for the population to fall from epoch $n$ (say) down to epoch
$n-1$. This happens when, due to fluctuations, all individuals of
fitness $n$ are lost from the population. We refer to this as a
destabilization of epoch $n$.  For some parameter settings, such
as shown in Figs. \ref{fig_runs}(a) and \ref{fig_runs}(c), this is
very rare. The time for the GA to reach the global optimum is mainly
determined by the time it takes to discover strings of fitness $n+1$
in each epoch $n$. For other parameter settings, however, such as in
Figs. \ref{fig_runs}(b) and \ref{fig_runs}(d), the destabilizations
play an important role in how the GA reaches the global optimum.
In these regimes, destabilization must be taken into account in
calculating search times. This task is accomplished in the
sequel \cite{Nimw98b}.

\subsection{Selection Operator}

We now explicitly construct the generation operator ${\bf G}$ for the
limit of infinite population size by constructing the selection
operator ${\bf S}$ and mutation operator ${\bf M}$. First, we
consider the effect of selection on the fitness distribution.
Since we are using fitness-proportionate selection, the
proportion $P^s_i$ of strings with fitness $i$ after selection is
proportional to $i$ and to the fraction $P_i$ of strings with
fitness $i$ before selection; that is,
\begin{equation}
P^s_i = c \; i \; P_i ~.
\end{equation}
The constant $c$ can be determined by demanding that the distribution
remains normalized:
\begin{equation}
1 =\sum_{i=1}^{n} P^s_i = c \sum_{i=1}^{N+1} i P_i ~.
\end{equation}
Since the average fitness $\langle f \rangle$ of the population is
given by:
\begin{equation}
\langle f \rangle = \sum_{i=1}^{N+1} i P_i ~,
\label{AverageFitness}
\end{equation}
we have
\begin{equation}
P^s_i = \frac{i P_i}{\langle f \rangle} ~.
\end{equation}
We can thus define a (diagonal) operator ${\bf S}$ that works on a
fitness distribution $\vec{P}$ and produces the fitness distribution
$\vec{P}^s$ after selection by:
\begin{equation}
\left({\bf S} \cdot \vec{P}\right)_i = \sum_{j=1}^{N+1} \frac{ \delta_{ij}
  j}{\langle f \rangle} P_j ~.
\end{equation}
Notice that this operator is nonlinear since, by Eq.
(\ref{AverageFitness}), the average fitness $\langle f\rangle$ is a
function of the distribution $\vec{P}$ on which the operator acts.

\subsection{Mutation Operator}
\label{sec_mut_operator}

The precise form of the mutation operator ${\bf M}$ depends on the
value $n$ of the current best fitness in the population. Thus, we
have to construct the mutation operator ${\bf M}$ for each possible
$n$. Assuming that strings of fitness
$n$ are the highest fitness strings in the current population we
calculate the probabilities ${\bf M}_{ij}$ that a string of fitness
$j$ is turned into a string with fitness $i$ under mutation, where the
indices $i$ and $j$ run from $1$ to $n$. Notice that, in this way, we
explicitly calculate the elements ${\bf M}_{ij}$ restricted to the
$n$-dimensional subspace of the fitness distribution space.

First, consider the components ${\bf M}_{ij}$ with $i < j$. These
strings are obtained if mutation leaves the first $i-1$ blocks of
the string unaltered and disrupts the $i$th block in the string. The
effect of mutation on the blocks $i+1$ through $N$ is immaterial
for this transition. Multiplying the probabilities that the preceding
$i-1$ blocks remain aligned and that the $i$th block becomes unaligned
we have:
\begin{equation}
{\bf M}_{ij} = (1-q)^{(i-1)K} \left( 1 - (1-q)^K\right), \;  \; i < j ~.
\end{equation}

The diagonal components ${\bf M}_{jj}$ are obtained when mutation
leaves the first $j-1$ blocks unaltered and does {\em not} mutate
the $j$th block to be aligned. The maximum entropy assumption says
that the $j$th block is random and so the probability $P_a$ that
mutation will change the unaligned $j$th block to an aligned block is
given by:
\begin{equation}
P_a = \frac{1-(1-q)^K}{2^K-1} ~.
\end{equation}
This is the probability that at least one mutation will occur in
the block times the probability that the mutated block will be in the
correct configuration. Thus the diagonal components are given by:
\begin{equation}
{\bf M}_{jj} = (1-q)^{(j-1)K} \left(1 -
  \frac{1-(1-q)^K}{2^K-1}\right).
\label{DiagonalMutation}
\end{equation}

Finally, we calculate the probabilities for increasing-fitness
transitions ${\bf M}_{ij}$ with $i > j$. These transition probabilities
depend on the states of blocks $j$ through $n-1$. One approximation,
using the maximum entropy assumption, is obtained by assuming that all
these blocks are random. The $j$th block is equally likely to be in any
of $2^K-1$ unaligned configurations. All succeeding blocks are equally
likely to be in any one of the $2^K$ configurations, including the
aligned one. In order for a transition to occur from state $j$
to $i$, first of all the $j-1$ aligned blocks have to remain
aligned, then the $j$th block has to become aligned through the
mutation. The latter has probability $P_a$. Furthermore, the following
$i-j-1$ blocks all have to be aligned. Finally, the $i$th block has to
be unaligned. Putting these together, we find that:
\begin{eqnarray}
{\bf M}_{ij} & = & (1-q)^{(j-1)K} \left( \frac{1-(1-q)^K}{2^K-1}\right)
  \nonumber \\
  & & \left( \frac{1}{2^K}\right)^{i-j-1}
  \left( 1 - \frac{1}{2^K} \right) ~, ~ i > j ~.
\end{eqnarray}
As a technical aside, note that for the case $i=n$ the last factor does
not appear. Since the current highest fitness in the population is $n$,
it is almost certain that the $n$th block is unaligned in all strings
in the population. As we show below in section \ref{genealogy_sec}, the
reason for this is that all individuals in the population during
epoch $n$ are descendants of strings in the highest fitness class $n$.
The strings in the highest fitness class $n$ have their $n$th block
unaligned by definition. If any string with fitness $i < n$ has the
$n$th block aligned, this block must have become aligned in the few
number of generations after it's appearance from a string of fitness
$n$. Generally, this only occurs with very low probability and so
can be ignored.

Another approximation to the components ${\bf M}_{ij}$ with $i > j$ is
obtained by assuming that all individuals with fitness $j$, for every
$j < n$, are offspring of an individual with fitness $n$ that had its
$j$th block become unaligned through mutation. This means that a
string of fitness $j$ has $n-2$ aligned blocks and one unaligned
block, namely, the $j$th block. Mutations from $j$ to $i$ with $i > j$
then only occur from $j$ to $i=n$ and do so by aligning the $j$th block:
\begin{equation}
{\bf M}_{ij} = \delta_{in} (1-q)^{(n-2)K} \frac{1 - (1-q)^K}{2^K-1} ~.
\end{equation}

It turns out that both approximations give very similar
results for observables such as epoch duration and total number of
fitness evaluations to reach the global optimum. In fact, to a good
approximation we can set all components ${\bf M}_{ij}$ with $ i > j$
to zero, since these components involve the alignment of at least one
block through mutation. For $K$ not too small this occurs with much
smaller probability than block destruction. That is, to a
good approximation, we can neglect terms proportional to $P_a$ in the
components ${\bf M}_{ij}$. 

The restricted generation operator ${\bf G}^n$ is now obtained by
taking the product of the selection operator ${\bf S}$ with the
mutation operator ${\bf M}$:
\begin{equation}
{\bf G}^n = {\bf M} \cdot {\bf S},
\end{equation}
where the component indices of the mutation and selection operators
run from $1$ to $n$.

\section{Quasispecies Distributions and Epoch Fitness Levels}
\label{an_appr}

During epoch $n$ the quasispecies fitness distribution $\vec{P}^n$
is given by a fixed point of the operator ${\bf G}^n$. To obtain this
fixed point we linearize the generation operator by taking out the
factor $\langle f \rangle$, thereby defining a new operator
${\bf \tilde{G}}^n$ via:
\begin{equation}
{\bf G}^n = \frac{1}{\langle f \rangle} {\bf \tilde{G}}^n.
\end{equation}
The operator ${\bf \tilde{G}}^n$ is just an ordinary (linear) matrix
operator and the quasispecies fitness distribution is nothing other
than the principal eigenvector of this matrix. The principal eigenvalue
$f_n$ of ${\bf \tilde{G}}^n$ is the average fitness of the quasispecies
distribution. In this way, obtaining the quasispecies distribution
reduces to calculating the principal eigenvector of the matrix
${\bf \tilde{G}}^n$. Again the reader is referred to Ref.
\cite{Nimw97b}.

As in Refs. \cite{Nimw97a} and \cite{Nimw97b}, the local stability of
the epochs can be analyzed by calculating the eigenvalues and
eigenvectors of the Jacobian matrix ${\bf DG}^n$ around each epoch
fitness distribution $\vec{P}^n$. The components ${\bf DG}^n_{ij}$ of
the Jacobian around epoch $n$ are given by:
\begin{equation}
{\bf DG}^n_{ij} = \left[ \frac{ \partial {\bf G}_i(\vec{P})}{\partial
    P_j}\right]_{\vec{P} = \vec{P}^n} = \frac{{\bf \tilde{G}}^n_{ij} -
    j \vec{P}^n_j}{f_n} ~.
\end{equation}
Just as in Ref. \cite{Nimw97b}, the eigenvectors $\vec{U}^i$ of the
Jacobian are given by $\vec{U}^i = \vec{P}^i - \vec{P}^n$, with
corresponding eigenvalues $e^n_i = f_i/f_n$. Thus, the spectra of the
Jacobian matrices are simply determined by the spectrum of the
generation operator ${\bf \tilde{G}}$. The eigenvalues $e^n_i$
determine the bulk of the GA's behavior. Since the matrix
${\bf \tilde{G}}$ is generally of modest size, i.e. its dimension is
determined by the number of blocks $N$, we can easily obtain numerical
solutions for the epoch fitnesses $f_n$ and the epoch quasispecies
distributions $\vec{P}^n$. At the same time one also obtains the
eigenvalues $e^n_i$ and eigenvectors $\vec{U}^i$ of the Jacobian.

For a clearer understanding of the functional dependence of the epoch
fitness distributions on the GA's parameters, however, we will now
develop analytical approximations to the epoch fitness levels $f_n$ and
quasispecies distributions $\vec{P}^n$.

In order to explicitly determine the form of the quasispecies
distribution $\vec{P}^n$ during epoch $n$ we must approximate the
matrix ${\bf \tilde{G}}^n$. As we saw in section \ref{sec_mut_operator},
the components ${\bf M}_{ij}$ (and so of $\bf G$) naturally fall
into three categories. Those with $i < j$, those with $i > j$, and
those on the diagonal $i=j$. Components with $i > j$ involve at least
one block becoming aligned through mutation. As noted above, these
terms are generally much smaller than the terms that only involve the
destruction of aligned
blocks or for which there is no change in the blocks. We therefore
approximate ${\bf \tilde{G}}^n$ by neglecting terms proportional
to $P_a$. Under this approximation for the components of
${\bf \tilde{G}}^n$, we have:
\begin{equation}
{\bf \tilde{G}}^n_{ij} = j (1-q)^{(i-1)K} (1 - (1-q)^K) , \; \; i < j ~,
\end{equation}
and 
\begin{equation}
{\bf \tilde{G}}^n_{jj} = j (1-q)^{(j-1) K} ~.
\end{equation}
The components with $i > j$ are now all zero. The result is that
${\bf \tilde{G}}^n$ is upper triangular. As is well known in
general matrix theory, the eigenvalues of an upper triangular matrix
are given by its diagonal components. Therefore, the average fitness
$f_n$ in epoch $n$, which is given by the largest eigenvalue, is equal
to the largest diagonal component ${\bf \tilde{G}}^n$. That is,
\begin{equation}
\label{fn_expression}
f_n = n(1-q)^{(n-1)K} ~.
\end{equation}
Notice that the matrix elements only depend on $q$ and $K$ through the
effective parameter $\lambda$ defined by:
\begin{equation}
\lambda = (1-q)^K ~.
\end{equation}
$\lambda$ is just the probability that a block will not be mutated.

The principal eigenvector $\vec{P}^n$ is the solution of the
equation:
\begin{equation}
\sum_{j=1}^n \left( {\bf \tilde{G}}^n_{ij} - f_n \delta_{ij}\right) P^n_j = 0 ~.
\end{equation}
Since the components of ${\bf \tilde{G}}^n$ depend on $\lambda$ in
such a direct way, we can analytically solve for this eigenvector;
finding that the quasispecies components are given by:
\begin{equation}
P^n_i = \frac{(1-\lambda) n \lambda^{n-1-i}}{n \lambda^{n-1-i} - i}
\prod_{j=1}^{i-1} \frac{n \lambda^{n-j} - j}{n \lambda^{n-1-j} -j} ~.
\end{equation} 
For the component $P^n_n$ this reduces to
\begin{equation}
P^n_n =
\prod_{j=1}^{n-1} \frac{n \lambda^{n-j} - j}{n \lambda^{n-1-j} -j} ~.
\label{pnn_expression}
\end{equation}
The above equation can be re-expressed in terms of the epoch fitness
levels $f_j$:
\begin{equation}
P^n_n =
\lambda^{n-1} \prod_{j=1}^{n-1} \frac{f_n- f_j}{f_n - \lambda f_j} ~.
\end{equation}

Note that it is straightforward to increase the accuracy of our
analytical approximations by including terms proportional to $P_a$ in
the matrix ${\bf \tilde{G}}^n$ and then treating these terms as a
perturbation to the upper triangular matrix. Using standard
perturbation theory, one can then obtain corrections due to block
alignments. We will not pursue this here, however, since the current
approximation is accurate enough for the optimization analysis.

\section{Quasispecies Genealogies and Crossover's Role}
\label{genealogy_sec}

Before continuing on to solve this problem, we digress slightly at this
point for two reasons. First, we claimed in a previous section that
{\em all} individuals in the population during epoch $n$ are descendants
of strings with fitness $n$. We will demonstrate this now by considering
the genealogies of strings occurring in a quasispecies population
$\vec{P}^n$. Second, since the argument is quite general and only
depends on the effects of selection, it has important implications for
population structure in metastable states (such as fitness epochs)
and, more generally, for the role of crossover in evolutionary search. 

For the $n$th epoch, let the set of ``suboptimal'' strings be all
those with fitness lower than $n$; their proportion
is simply $1-P^n_n$. This proportion is constant on
average during epoch $n$. Over one generation, the suboptimal strings
in the next generation will be either descendants of suboptimal
strings in the current generation or mutant descendants of
``optimal'' strings with fitness $n$. Let $r$ denote the average
number of offspring per suboptimal individual. The fact that the total
proportion of suboptimal strings remains constant gives us the
equation:
\begin{equation}
(1-P^n_n) = r (1-P^n_n) + \frac{(1-{\bf M}_{nn})n}{f_n} P^n_n ~.
\end{equation}
The last term is the proportion of individuals of fitness $n$ that are
selected and do not stay at fitness $n$ under mutation. From this
we can solve for $r$ to find:
\begin{equation}
r = 1 - \frac{(1-{\bf M}_{nn}) n P^n_n}{f_n (1-P^n_n)} = \frac{1 -
  \lambda^{1-n} P^n_n}{1-P^n_n} ~,
\end{equation}
where we have used the previous analytical approximations to $f_n$ and
${\bf M}_{nn}$, Eqs. (\ref{fn_expression}) and (\ref{DiagonalMutation}),
in the last equality.

We see that in every generation only a fraction $r$ of the suboptimal
individuals are descendants from suboptimal individuals in the previous
generation. This means that after $t$ generations, only a fraction
$r^t$ of the suboptimal individuals are the terminations of lineages
that solely consist of suboptimal strings. There is a fraction of
$1-r^t$ strings that have one or more ancestors with fitness $n$ in
the $t$ preceding generations. After a certain number of generations
$t_c$ this fraction becomes so small that less than one individual
(on average) has only suboptimal ancestors. That is, after approximately
$t_c$ generations in epoch $n$, the whole quasispecies will consist of
strings that are descendants of a string with fitness $n$ somewhere in
the past.  Explicitly, we find that
\begin{equation}
t_c = \frac{\log \left[ M(1-P^n_n)\right]}{\log \left[ 
\frac{1-\lambda^{1-n} P^n_n}{1-P^n_n} \right]} ~.
\end{equation}
As expected $t_c$ is proportional to the logarithm of the total number
$M(1-P_n^n)$ of suboptimal strings in the quasispecies.

The above result can be generalized to the case in which the suboptimal
strings are defined to be all those with fitnesses $1$ to $n-i$. One
can then calculate the time until all strings in these classes are
descendants of strings with fitness $n-i+1$ to $n$ to find that the
lower classes are taken over even faster by descendants of strings
with fitness $n$.

The preceding result is significant for a conceptual understanding of
the structure of epoch populations. All suboptimal individuals in the
population have an ancestor of optimal fitness that is a relatively
small number of generations in the recent past. The result is that in
genotype space
all suboptimal individuals are always relatively close to some
individual of optimal fitness. The suboptimal individuals never
wander far from the individuals of optimal fitness. More precisely,
the average length of suboptimal lineages is $1/(1-r)$ generations.
That is, all suboptimal individuals are to be found within an average
Hamming distance of $L q/(1-r)$ from optimal-fitness individuals. The
individuals with optimal fitness, however, {\em can} wander through
genotype space as long as they do not leave the neutral network
of optimal fitness strings---those with fitness $n$ in epoch $n$.
If the population is to traverse large regions of genotype space in
order to discover a string of fitness larger than $n$, it must do so
along this neutral network. In short, this
is the reason we believe the existence of neutral networks, consisting
of approximately equal fitness strings that percolate through large
parts of the genotype space, is so important for evolutionary search;
cf. Ref. \cite{Huynen&Stadler&Fontana}. If strings of fitness $n$ were
to form a small island in a sea of much lower fitness strings that are
at relatively large Hamming distances from islands with fitness higher
than $n$, then there is little chance that a suboptimal fitness mutant
will drift far enough from the island of fitness $n$ strings to discover
another island with higher fitness.

This result---that all strings in the metastable population are relatively
recent descendants of strings in the highest fitness class---should
generalize to other selection schemes such as elite selection, rank
selection, and tournament selection. Furthermore, this view also
provides some insight into the effects of adding crossover to
evolutionary search algorithms. Assume that we add crossover to out
current GA; see also the discussion of similar crossover experiments
in section 6.5 of Ref. \cite{Nimw97b}.
The initial population still has a distribution
of fitnesses given by Eq. (\ref{init_fit_dist}). The best fitness in
the initial population might be (say) $3$, corresponding to the first
two blocks being aligned and the third block unaligned. It is unlikely
that crossovers will lead quickly to strings of fitness $4$. Although
the initial population will have strings with the $3$rd block aligned,
these strings are very unlikely to also have the first block aligned.
This means that these strings have low fitness and so are unlikely to
be selected as the parent of a crossover event. Moreover, relatively
quickly, the entire population will become descendants of strings with
fitness $3$ that, by definition, have the third block unaligned.
Crossover events will thus almost never lead to the creation of strings
of higher fitness; at least not through the ``combining of building
blocks'' \cite{Holland92a}.

The positive contribution of crossover is that an aligned block may be
formed from two parents each with an unaligned $3$rd block if the
crossover point falls within the $3$rd block and if the resulting
complementary subblocks are themselves aligned. The negative effect
is that crossover may also
combine lower fitness strings with higher fitness strings so as to
produce two lower fitness offspring. Thus, with nothing
else said or added, we expect the effect of
crossover used in the GA to be marginal. Experiments with single-point
crossover confirm this. The global optimum is found somewhat
more quickly, but the improvement in search time is very small and often
washed out by the large variance in search time. We return
to this issue in Ref. \cite{Nimw98b}.

These arguments are specific to the Royal Staircase (and also Royal
Road) classes of fitness function. However, for evolutionary dynamics
exhibiting epochal evolution we believe it to be the case that the
population structure is a cloud of strings {\em localized} around
strings of the current best fitness. Therefore, the effect of crossover
mostly will be to increase the amount of mixing {\em within} the
quasispecies cloud. That is, crossover acts during the epochs as a
local mixing operator very much as mutation does.

\section{Mutation Rate Optimization}
\label{sec_mut_rate_opt}

In the previous sections we argued that the GA's behavior can be viewed
as stochastically hopping from epoch to epoch until the search
discovers a string with fitness $N+1$. Assuming for the moment that
the total time to reach this global optimum is dominated by the time
the GA spends in the epochs, we will derive a way to tune the mutation
rate $q$ such that the time the GA spends in an epoch is minimized.

During epoch $n$ no string in the population has the $n$th block
aligned. Thus, the main contribution to the waiting time in epoch $n$ is
given by the time it takes a mutant with the $n$th block set correctly
to appear. As we have seen, the probability $P_a$ to
mutate a single block such that it becomes aligned is given by:
\begin{equation}
P_a = \frac{1 - (1-q)^K}{2^K -1} = \frac{1-\lambda}{2^K -1},
\end{equation}
where again $\lambda$ is the probability that a block will not be
mutated at all. Obviously, the higher $P_a$, the more likely mutation
is to produce a new aligned block. Every generation each individual
in the population has a probability $P_a$ of aligning the $n$th block.
Aligning the $n$th block only creates a string of fitness $n+1$ when
all the $n-1$ blocks to its left are aligned as well. That is, only
the fraction $P^n_n$ of the population with fitness $n$ will produce a
fitness $n+1$ string by aligning the $n$th block. Therefore, the
probability $C_{n+1}$ that a string of fitness $n+1$ will be created
over one generation is given by:
\begin{equation}
C_{n+1} \approx M P^n_n P_a ~.
\label{ProbToCreateString}
\end{equation}
Our claim is that, as a first approximation, maximizing $C_{n+1}$
minimizes the number of generations the population spends in epoch $n$. 

In section \ref{an_appr} we derived an analytic approximation to the
proportion $P^n_n$ of individuals in the highest fitness class during
epoch $n$ as a function of $\lambda = (1-q)^K$. Since $P_a$ is also
a monotonic function of $\lambda$, as it is of $q$, we can maximize
$C_{n+1}$ as a function of $\lambda$ instead. Ignoring proportionality
constants, the function to maximize is:
\begin{equation}
\label{simplest_opt_approx}
C_{n+1} \propto (1-\lambda) P^n_n(\lambda) ~.
\end{equation}
Using Eq. (\ref{pnn_expression}) for the dependence of $P^n_n$ on
$\lambda$ and differentiating the above function $C_{n+1}$ with
respect to $\lambda$, we find that the optimal $\lambda_o$
satisfies:
\begin{equation}
\frac{n (1-\lambda_o)}{\lambda_o} \left[ \sum_{i=1}^{n-1} \frac{i
    \lambda_o^{i}}{n \lambda_o^i - n +i} - \frac{(i-1) \lambda_o^{i-1}}{n
    \lambda_o^{i-1} -n + i}\right] = 1 ~.
\end{equation}
The solution of this equation gives the optimal $\lambda_o(n)$ which
is only a function of the epoch number $n$. This is an important
observation, since it means that the optimal value of the mutation
rate $q_o$ takes the following general form as a function of $n$ and
$K$:
\begin{equation}
q_o = 1 - \sqrt[K]{\lambda_o(n)} ~.
\label{OptimalMutation}
\end{equation}
Once we solve for the function $\lambda_o(n)$, we can immediately
obtain the dependence of $q_o$ on $K$ using Eq. (\ref{OptimalMutation}).

In this calculation we assumed that the waiting time for
{\em discovering} a higher fitness string dominated the time spent in
an epoch. This means that as soon as a string of fitness $n+1$ is
created, copies of this string spread through the population and
the population stabilizes onto epoch $n+1$. In fact, it is quite
likely that the string with fitness $n+1$ will be lost through a
deleterious mutation in the aligned blocks or via sampling before it
gets a chance to establish itself in the population. Recall the
transitory jumps in the best fitness seen in Fig. \ref{fig_runs}.
In Ref. \cite{Nimw97b} we used a diffusion equation approximation to
calculate the probability $\pi_n$ that a string with fitness $n+1$
will spread. We found that to a good approximation it is given by:
\begin{equation}
\label{pi_expression}
\pi_n = \frac{1 - \left( 1 - \frac{1}{M}\right)^{2 M \gamma_n +1}}{1-
  \left(1 - P^{n+1}_{n+1}\right)^{2 M \gamma_n+1}}  \approx 1- e^{-2 \gamma_n} ~,
\end{equation}
where
\begin{equation}
\gamma_n = \frac{f_{n+1} - f_n}{f_n},
\end{equation}
and the last approximation in Eq. (\ref{pi_expression}) holds for
relatively large population sizes. Notice that the spreading
probability $\pi_n$ only depends on the relative difference of the
average fitness in epoch $n+1$ and epoch $n$. Using Eq.
(\ref{fn_expression}) for $f_n$ we find:
\begin{equation}
\gamma_n = \left( 1 + \frac{1}{n}\right) \lambda - 1 ~.
\end{equation}
Thus, we find that $\pi_n(\lambda)$ is approximately given by:
\begin{equation}
\pi_n(\lambda) = 1 - e^{-2 \left( 1 + \frac{1}{n}\right) \lambda +2} ~.
\label{ProbToStabilize}
\end{equation}
The probability $C_{n+1}'$ to create a string of fitness $n+1$ that
spreads through the population is thus given by:
\begin{equation}
\label{corrected_opt_approx}
C_{n+1}' = C_{n+1} \pi_n(\lambda) ~.
\end{equation}
Taking the spreading probability $\pi_n$ into account, we want to
maximize $C_{n+1}'$ in order to minimize the time spent in epoch $n$.
Note that, also in this case, the dependence on $q$ and $K$ enters only
through $\lambda$. For each $n$ there an optimal value of $\lambda$
from which the optimal mutation rate can be determined as a function
of $K$.

\begin{figure}[htbp]
\centerline{\epsfig{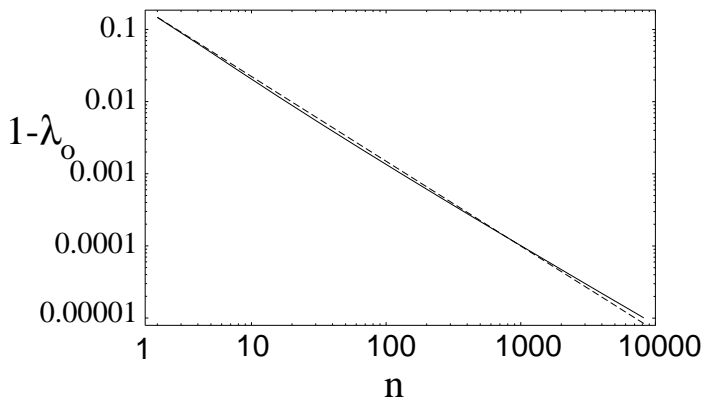}}
\caption{Optimal $\lambda$ as a function of $n$. The vertical axis
shows $1-\lambda_0$ on a logarithmic scale. The horizontal axis plots
$n$ on a logarithmic scale. The solid line is the numerical
solution of Eq. (\ref{eq_lam_opt2}) and the dashed line is the
approximation given by Eq. (\ref{approx_lam_opt2}).}
\label{lam_opt2}
\end{figure}

The optimal value $\lambda_o$ in this case is the solution of:
\begin{eqnarray}
\label{eq_lam_opt2}
\frac{n (1-\lambda_o)}{\lambda_o} & & \left[ \sum_{i=1}^{n-1} \frac{i
    \lambda_o^{i}}{n \lambda_o^i - n +i} - \frac{(i-1) \lambda_o^{i-1}}{n
    \lambda_o^{i-1} -n + i}\right] \nonumber \\
	& + & (1-\lambda_o) \frac{2
    \left(1+\frac{1}{n}\right) e^{-2 \left(1 + \frac{1}{n}\right)
    \lambda_o +2}}{1 - e^{-2 \left(1 + \frac{1}{n}\right)
    \lambda_o +2}} = 1.
\end{eqnarray}
Numerically, the solution $\lambda_o(n)$ is well approximated by:
\begin{equation}
\label{approx_lam_opt2}
\lambda_o(n) = 1 - \frac{1}{3 n^{1.175}} ~,
\end{equation}
as shown in Fig. \ref{lam_opt2}, which plots $(1-\lambda_o)$ as a
function of $n$. The solid line is the numerical solution obtained
from Eq. (\ref{eq_lam_opt2}); the dashed line is the approximation
Eq. (\ref{approx_lam_opt2}).

For large $n$, using Eq. (\ref{approx_lam_opt2}) we can approximate
the optimal mutation rate by:
\begin{equation}
q_o = \frac{1}{3 (n K) n^{0.175}} ~.
\end{equation}
Thus, the optimal mutation rate drops as a power-law in both $n$ and
$K$. This implies that, generally for the Royal Staircase fitness
function, the mutation rate should decrease as a GA run progresses
so that the search will find the global optimum as quickly as
possible. We pursue this idea more precisely elsewhere by considering
an adaptive mutation rate scheme for the GA.

\section{Fitness Function Evaluations: Theory versus Experiment}

In the preceding sections we derived an expression for the probability
$C_{n+1}'$ to create, over one generation in epoch $n$, a string of
fitness $n+1$
that will stabilize by spreading through the population. From this we
now estimate the total number $E$ of fitness function evaluations
the GA uses on average before an optimal string of fitness $N+1$ is
found. As a first approximation, we assume that the GA visits all
epochs, that the time spent in innovations between them is negligible,
and that epochs are always {\em stable}. By epoch stability we mean it
is highly unlikely that strings with the current highest fitness will
disappear from the population through a fluctuation, once such strings
begin to spread. These assumptions appear to hold for
the parameters of Figs. \ref{fig_runs}(a) and \ref{fig_runs}(c). They
may hold even for the parameters of Fig. \ref{fig_runs}(b), but
they probably do not for Fig. \ref{fig_runs}(d). For the parameters
of Fig. \ref{fig_runs}(d), we see that the later epochs ($n = 8$, $9$,
and $10$) easily destabilize a number of times before the global
optimum is found. We will develop a generalization that addresses this
more complicated behavior in Ref. \cite{Nimw98b}.

The average number $T_n$ of generations that the population spends in
epoch $n$ is simply $1/C_{n+1}'$, the inverse of the probability that
a string of fitness $n+1$ will be discovered and spread through the
population. For a population of size $M$, the number of fitness
function evaluations per generation is $M$, so that the total number
$E_n$ of fitness function evaluations in epoch $n$ is given by $T_n M$.
More explicitly, we have from this and Eqs. (\ref{ProbToCreateString})
and (\ref{corrected_opt_approx}) that:
\begin{equation}
E_n = \frac{2^K-1}{(1-\lambda)  P^n_n \pi_n} ~.
\end{equation}
This says that, given our approximations, the total number of fitness
function evaluations in each epoch is independent of the population
size $M$. The epoch lengths, measured in generations, are inversely
proportional to $M$, while the number of fitness function evaluations
per generation is $M$. Substituting our analytical expressions
for $P^n_n$ and $\pi_n$, Eqs. (\ref{pnn_expression}) and
(\ref{ProbToStabilize}), we have:
\begin{equation}
E_n(\lambda) = \frac{2^K-1}{(1-\lambda)
  \left(1-e^{-2\left(1+\frac{1}{n}\right) \lambda +2}\right)}
\prod_{i=1}^{n-1} \frac{n \lambda^{n-i-1} -i}{n \lambda^{n-i} -i} ~.
\label{approx_En}
\end{equation}
In Ref. \cite{Nimw98b} we use this to derive analytical scaling
expressions for the minimal number of function evaluations that the
GA requires on average in epoch $n$ as given by $E_n(\lambda_o)$,
where $\lambda_o$ is the optimal $\lambda$ for epoch $n$.

The total number of fitness function evaluations $E(\lambda)$ to reach
the global optimum is given by the sum of $E_n(\lambda)$
over all epochs $n$ from $1$ to $N$:
\begin{equation}
\label{tot_lam}
E(\lambda) = \sum_{n=1}^{N} \frac{2^K-1}{(1-\lambda) \pi_n(\lambda)} \prod_{i=1}^{n-1}
    \frac{n \lambda^{n-i-1} -i}{n \lambda^{n-i} -i} ~.
\end{equation}
The optimal mutation rate for an entire run is obtained by
minimizing the above expression for $E$ with respect to $\lambda$.

\begin{figure}[htbp]
\centerline{\epsfig{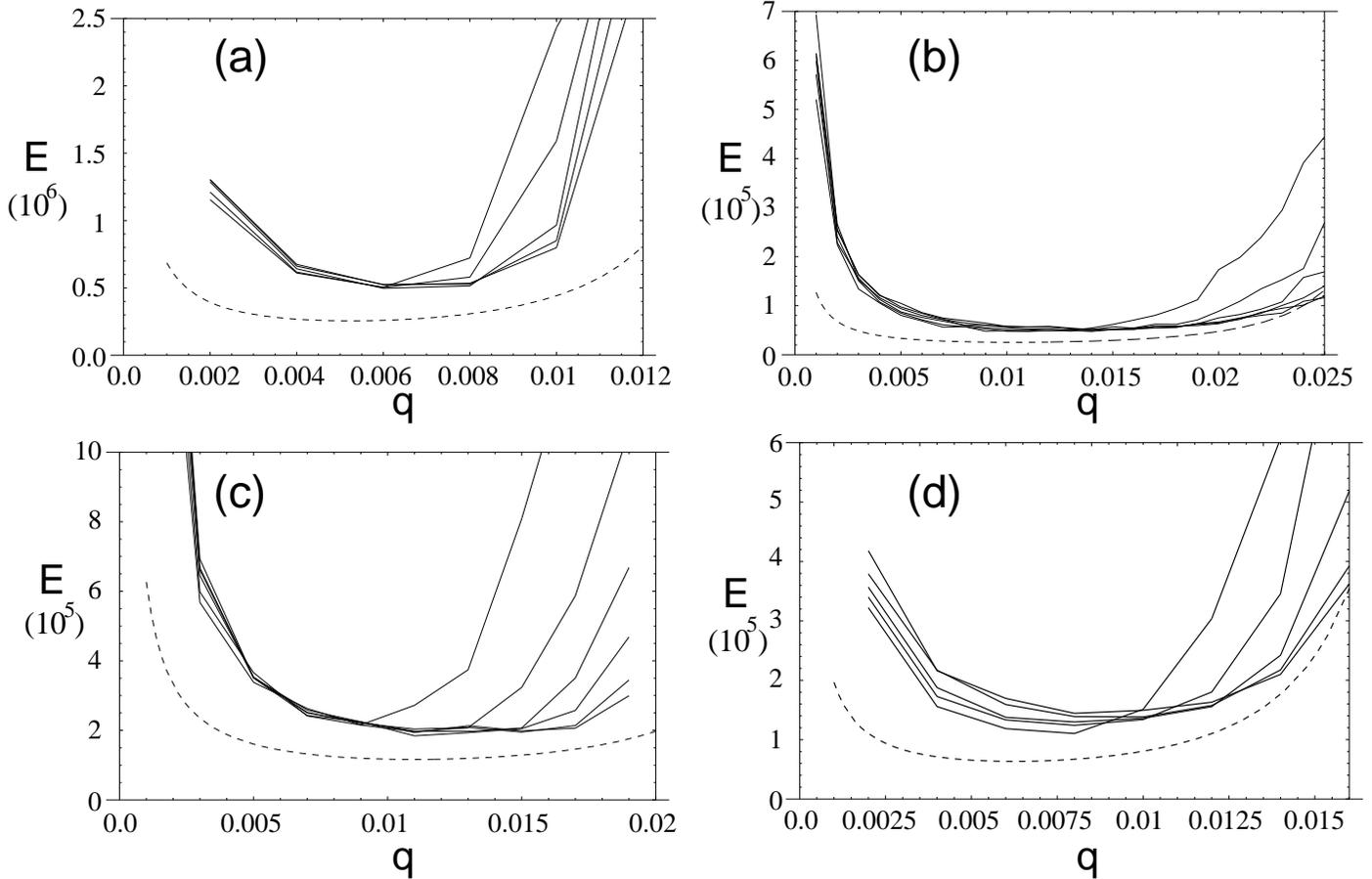}}
\caption{Comparison of experimental results (solid lines) and
  theoretical predictions (dashed lines) of the total number $E$
  of function evaluations to reach the global
  optimal as a function of the mutation rate $q$, for four different
  fitness functions as determined by $N$ and $K$. The parameters are
  the same as those used in Fig. \ref{fig_runs}. Plot (a) has $N=8$
  and $K=8$, (b) has $N=6$ and $K=6$, (c) has $N=4$ and $K=10$, and
  (d) has $N=10$ and $K=5$. The dashed lines give the theoretical
  predictions according to Eq. (\ref{tot_lam}). The solid lines are the
  results of experiments with different population sizes $M$. Each data
  point on the solid lines gives $E$ averaged over $R =250$ GA runs.
  Each solid line shows the estimated $E$ as a function of
  $q$ for a constant population size $M$. For low mutation rates, the
  solid lines approximately overlap, indicating that $E$ is
  approximately independent of $M$ in this regime. For large mutation
  rates, the lower values of $M$ have larger $E$ than for larger $M$.
  Plot (a) shows, from left to right, population sizes $M=150$, $M=200$,
  $M=250$, $M=300$, and $M=350$; plot (b) $M=60$, $M=90$, $M=120$,
  $M=150$, $M=180$, and $M=210$;  plot (c) $M=50$, $M=80$, $M=110$,
  $M=140$, $M=170$, and $M=200$; and finally, plot (d) $M=100$,
  $M=150$, $M=200$, $M=250$, and $M=300$.}
\label{comp_mosaic}
\end{figure}

Figure \ref{comp_mosaic} compares Eq. (\ref{tot_lam}) to the number
(solid lines) of function evaluations estimated from 250 GA runs
for the four different settings of $N$ and $K$ of Fig. \ref{fig_runs}.
Each plot is a function of mutation rate $q$ and is given for a set
of different population sizes $M$.

\end{multicols}

\begin{multicols}{2}

Each of the four plots in Fig. \ref{comp_mosaic} shows, as a dashed
line, the population-size independent theoretical approximation of the
total number $E$ of fitness function evaluations as a function of
$q$. The four parameter settings of $N$ and $K$ are the same as in the
runs (a), (b), (c), and (d) of Fig. \ref{fig_runs}. In each plot in Fig.
\ref{comp_mosaic} the solid lines show $E$ as a function of $q$ for 
several different population sizes as obtained from GA experiments.
Each data point on the solid lines is an average over $250$ GA runs.

Figure \ref{comp_mosaic}(a) has $N=8$ blocks of $K=8$ bits. The
dependence of $E$ on mutation rate is shown over the range from
$q=0.001$ to $q=0.012$.
The optimal mutation rate occurs somewhere around $q_o = 0.006$.  Each
solid line shows the experimental dependence of $E$ on $q$ for a fixed
population size. At high mutation rates, the lower population sizes
have higher $E$. At high mutation rates, the set of solid lines in
Fig. \ref{comp_mosaic} show, from left to right, $E(q)$ for population
sizes $M=150$, $M=200$, $M=250$, $M=300$, and $M=350$, respectively.

Figure \ref{comp_mosaic}(b) has parameters $N=6$ and $K=6$. $E(q)$ is
shown over the range $q=0.001$ to $q=0.025$. Again, at large
mutation rates the smaller population sizes have higher $E(q)$. The
solid lines from top to bottom in the high mutation rate regime show
$E(q)$ for population sizes $M=60$, $M=90$, $M=120$, $M=150$, $M=180$,
and $M=210$, respectively. The optimal mutation rate occurs somewhere
around $q_o = 0.012$.

Figure \ref{comp_mosaic}(c) has $N=4$ blocks of
length $K=10$ as parameter settings. $E(q)$ is shown over the range
$q=0.001$ to $q=0.02$, the optimum occurring around $q_o = 0.011$. The
solid lines shows the experimental $E(q)$ for population sizes $M=50$,
$M=80$, $M=110$, $M=140$, and $M=200$.

Finally, Fig. \ref{comp_mosaic}(d) has $N=10$ blocks of length $K=5$
bits. The horizontal axis ranges from $q = 0.001$ to $q = 0.016$ with
the optimal mutation rate occurring around $q_o = 0.008$.
The population sizes are, from top to bottom at high $q$, $M=100$,
$M=150$, $M=200$, $M=250$, and $M=300$, respectively. Note that
for figures \ref{comp_mosaic}(b), \ref{comp_mosaic}(c), and
\ref{comp_mosaic}(d) the tick marks on the vertical axis have a scale
of $10^5$ fitness function evaluations, while Fig. \ref{comp_mosaic}(a)
uses a scale of $10^6$ fitness function evaluations.

Note that in obtaining the theoretical predictions we have set
$\pi_N = 1$ in Eq. (\ref{tot_lam}), since by definition the GA stops
at the {\em first} occurrence of the globally optimum string. For the
last epoch $N$ it is only necessary to create a string of fitness $N+1$,
it does not need to spread through the population.

Figure \ref{comp_mosaic} shows that for all these different
parameter settings,\footnote{Each case has strings of comparable
numbers of blocks, between $N=4$ and $N=10$, and block lengths,
between $K=5$ and $K=10$. This is mainly done since it is
computationally expensive, if not infeasible, to obtain extensive
experimental data for substantially larger values of $N$ and $K$.}
the theory, which is independent of population size $M$, reasonably
predicts the location of the optimal mutation rate $q_o$. It also
predicts moderately well the shape of the curve around the optimum.

It is notable that the theory consistently underestimates $E(q)$.
Furthermore, it underestimates $E(q)$ more in the low mutation rate
regime than in the high. We believe that both of these offsets can
be explained by the effects of finite-population sampling.
We assumed that all unaligned blocks in members of the current highest
fitness class are random and {\em independent} of each other. This last
assumption does not hold in general \cite{Derrida&Peliti}. Due to
finite-population sampling and the resulting tendency to fixate,
strings in the highest fitness class are correlated with each other.
This means that if one individual in the highest fitness class has it's
$n$th unaligned block at a large Hamming distance from the desired
configuration (with that block aligned), then many individuals (as
descendants) in the highest fitness class also tend to have their $n$th
blocks at large Hamming distances from the desired configuration. This
genetic correlation among the individuals increases the epoch durations.
Moreover, this effect is more severe for small mutation rates which,
along with small population sizes, increase population convergence.

It turns out that this effect is difficult to analyze quantitatively. In
spite of this it appears that, for low mutation rates on up to the optimal
mutation rate, the total number $E$ of fitness function evaluations is
indeed approximately independent of population size $M$. Moreover,
the theory still accurately predicts the location of the optimal mutation
rate $q_o$. Also, although the exact magnitude of $E$ is underestimated,
the largest deviation in $E$ from the experimental minimum is $47$\%
for the parameters of Fig. \ref{comp_mosaic}(a), whereas the minimal
deviation is $37$\%, for the parameters of Fig. \ref{comp_mosaic}(c).
Thus, the theory predicts the correct order of magnitude for the minimal
number of fitness function evaluations.

As was noted above, the experimental curves for different population
sizes appear to collapse onto each other in the low mutation rate
regime. As the mutation rate increases, the solid curves break off this
common curve one by one and do so delayed in proportion to population
size. As the mutation rate increases it appears that progressively
larger population sizes are necessary to maintain the search's
efficiency. Of course, this effect is not explained by our
population-size independent theory. It is the topic of the sequel
\cite{Nimw98b}.

From the point of view of GA practice, it is important to emphasize
again that there is a large variance between GA runs (at fixed parameters)
in the total number of function evaluations to reach the global optimum.
For example, if there is an average number of $10^5$ fitness function
evaluations, then the standard deviation of the total number of fitness
function evaluations is also almost $10^5$. Some runs may take as much as
$5 \times 10^5$ fitness function evaluations (say) and some may only take
$10^4$ evaluations. Therefore, if one intends to use a GA for only
a few runs, or maybe even just once, there is a large range of $q$ and $M$
for which the performance of the GA looks statistically identical.
In other words, the GA has a large ``sweet spot'' with respect to
parameter settings for optimal search. Note that the optimal mutation rate
is generally well below $q = 1/L$. Thus, despite the large parameter sweet
spot, a mutation rate of $q = 1/L$ is far too large for epochal evolutionary
search.

\section{Conclusion and Future Analyses}

In summary our findings are the following.
\begin{enumerate}
\item{At fixed population size $M$, there is a smooth cost surface
	$E(q)$ as a function of mutation rate $q$. It has a {\em single}
    and {\em shallow} minimum $q_o$.}
\item{The optimal mutation rate $q_o$ roughly occurs in the regime
    where the highest epochs $N-1$, $N$, and $N+1$ are marginally
	stable; see Fig. \ref{fig_runs}.}
\item{For lower mutation rates than $q_o$ the total number of fitness
    function evaluations $E(q)$ grows steadily and becomes almost
    independent of the population size $M$.}
\item{Crossover's role in reducing search time is marginal
	due to population convergence during the epochs.}
\item{There is an mutational error threshold in $q$ that bounds the
    upper limit of the GA's efficient search regime. Above the
	threshold, which is population-size independent, suboptimal
	strings of fitness $N$ cannot stabilize in the population.}
\item{The surface $E(q)$ appears to be everywhere concave. That
    is, for any two parameters $q_1$ and $q_2$, the straight
    line connecting these two points is everywhere above the surface
    $E(q)$. We conjecture that this is always the case for mutation-only
	genetic algorithms with a static fitness function. This feature is
	useful in that a steepest-descent algorithm applied to parameter
	$q$ will lead to the unique optimum $q_o$.}
\end{enumerate}
Synopsizing the results this way, we are anticipating some of the
results developed for the population-size dependent theory
\cite{Nimw98b}.

In this sequel we extend the above statistical dynamics analysis to
account for $E(q)$'s dependence on population size. This not only
improves the parameter-optimization theory, but also leads us to
consider a number of issues and mechanisms that shed additional
light on how GAs work and on those types of search problems (fitness
functions) for which evolutionary search is and is not well suited.
Since it appears that optimal parameter settings often lead the GA
to run in a mode were the population dynamics is marginally stable
in the higher epochs, we consider how epoch destabilization affects
epoch stability and duration. We also analyze an adaptive evolutionary
search algorithm in which the mutation rate and population size change
dynamically as successive epochs are encountered. This leads to a
reduction in search time and in resource requirements.

\section*{Acknowledgments}

The authors thank Melanie Mitchell for helpful discussions. This work
was supported at the Santa Fe Institute by NSF Grant IRI-9705830, ONR
grant N00014-95-1-0524 and by Sandia National Laboratory contract
AU-4978.

\bibliography{epev} 
\bibliographystyle{plain}

\end{multicols}

\end{document}